\documentclass[10pt,conference]{IEEEtran}
\IEEEoverridecommandlockouts

\usepackage{cite}
\usepackage{amsmath,amssymb,amsfonts}
\usepackage{algorithmic}
\usepackage{graphicx}
\usepackage{textcomp}
\usepackage{xcolor}
\usepackage{multirow}
\usepackage{pifont}
\usepackage{sepnum}
\usepackage{xspace}
\usepackage[group-separator={,}]{siunitx}
\usepackage[most]{tcolorbox}
\usepackage{subcaption}
\usepackage{bigstrut}
\usepackage{makecell}
\usepackage{booktabs}
\usepackage{enumitem}
\usepackage{wrapfig}
\usepackage{dirtytalk}
\usepackage{colortbl}
\usepackage{xurl}
\usepackage{hyperref}
\usepackage{cleveref}

\definecolor{gainsboro}{rgb}{0.86, 0.86, 0.86}
\newcommand{\clg}[1]{\cellcolor{gainsboro}{#1}}
\newcommand{\xmark}{\ding{55}}%

\newcommand{\point}[1]{\par\smallskip\noindent\textbf{#1.} }
\newcommand{\fvitem}[1]{\par\textit{#1.}\ }
\newcommand{\rqref}[1]{\hyperlink{rq:#1}{RQ#1}}

\newtcolorbox{takeawaybox}[1][]{
    breakable,
    enhanced,
    colback=gray!4,
    colframe=black!65,
    boxrule=0.6pt,
    arc=1mm,
    left=4pt,
    right=4pt,
    top=4pt,
    bottom=4pt,
    before skip=6pt,
    after skip=6pt,
    #1
}

\newcommand{\circomspect}{{\sc Circomspect}\xspace}
\newcommand{\ecne}{{\sc Ecne}\xspace}
\newcommand{\picus}{{\sc Picus}\xspace}
\newcommand{\civer}{{\sc Civer}\xspace}
\newcommand{\zkfuzz}{{\sc zkFuzz}\xspace}

\newcommand{\acfour}{{\sc AC4}\xspace}
\newcommand{\ccccheck}{{\sc CCC-Check}\xspace}

\begin{document}

\title{SoK: ZKP Security Tools and Verification}

\author{%
\IEEEauthorblockN{Arman Kolozyan\textsuperscript{*}}
\IEEEauthorblockA{\textit{Max Planck Institute for}\\\textit{Security and Privacy (MPI-SP)}}
\and
\IEEEauthorblockN{Tom Sorger\textsuperscript{*}}
\IEEEauthorblockA{\textit{KTH}\\\textit{Royal Institute of Technology}}
\and
\IEEEauthorblockN{Alexander Hicks}
\IEEEauthorblockA{\textit{Ethereum Foundation}}
\and
\IEEEauthorblockN{Stefanos Chaliasos}
\IEEEauthorblockA{\textit{University of Athens \&}\\\textit{zkSecurity}}
\thanks{\textsuperscript{*}These authors have equally contributed to this work.}
}

\maketitle

\begin{abstract}
Zero-knowledge proofs (ZKPs) have become a core technology for privacy and verifiable computing. They are used to secure blockchains that handle billions of dollars and identity applications dealing with sensitive personal data. However, ZKP systems are complex, and subtle implementation errors can completely break their guarantees, letting attackers forge money or false proofs of identity. Researchers and practitioners have therefore developed a growing set of bug detection and formal verification methods to secure these systems. Yet their real-world effectiveness and adoption remain unclear.

In this paper, we aim to shed light on the state of ZKP security tooling. We first systematize the landscape of these tools and observe that most target Circom, leaving newer DSLs and zkVMs with limited support. We then evaluate six tools across 70 real-world vulnerabilities and find that while the tools detect 45.7\% of bugs on isolated targets, their effectiveness drops to 19.6\% on full codebases, with important vulnerability classes left unaddressed. We also present the first systematic analysis of formal verification efforts, revealing that current work focuses primarily on constraint correctness and identifying key gaps and risks. Finally, we survey 48 practitioners, showing that development and security remain human-led, LLMs are widely used, and practitioners prioritize tools with clearer guarantees and lower integration effort. 
Overall, our results highlight the need for better integration of security tooling with the development and auditing process, and we provide actionable insights for researchers and practitioners.
\end{abstract}


\section{Introduction}
Zero-knowledge proofs (ZKPs) are cryptographic protocols that let a prover convince a verifier that a statement is true, optionally without revealing anything beyond its validity~\cite{goldwasser1985knowledge}.
Over the last decade, advances in proof systems and implementations have made ZKPs practical, enabling emerging applications such as private payments~\cite{bensasson2014zerocash}, identity protocols~\cite{baldimtsi2024zklogin}, blockchain scaling through zk-rollups~\cite{chaliasos2024analyzing}, and verifiable machine learning~\cite{peng2025survey}.

These applications rely on strong properties such as succinct verification, privacy, and soundness~\cite{groth2016pairing}. 
Yet ZKP implementations are difficult to get right, as subtle bugs in circuits, proof systems, or integrations can break their guarantees, allowing false statements to get accepted by the verifier~\cite{chaliasos2024sok}.
These risks are not hypothetical. The Orchard vulnerability in Zcash remained hidden for years and could have enabled counterfeiting in its largest shielded pool~\cite{wilcox2026orchard}, an Aztec Connect exploit drained roughly \$2.1 million~\cite{aztec2026connect}, and high-severity ZKP bugs continue to appear in bug-bounty programs~\cite{zksync2026airbenderbounty}.

In response, researchers and practitioners have developed a growing set of defenses for ZKP implementations, including static analyzers~\cite{trailofbits2024circomspect,wen2024practical,soureshjani2023halo2,Schaeffer2023Pilspector,kolozyan2025dcm}, SMT-based verifiers~\cite{pailoor2023automated,liu2024certifying,yang2026ac4,stephens2025automated}, and fuzzers~\cite{takahashi2025zkfuzz,chaliasos2025fuzzing}. 
Given the complexity and security-critical role of ZKPs, formal verification is becoming increasingly important, with recent efforts targeting zero-knowledge virtual machines (zkVMs), proof systems, and verifier implementations~\cite{succinct2025sp1,openvm2026fv,kwan2026jolt,certik2024zkwasm,nethermind2025certiplonk}.
Nevertheless, the impact of these techniques remains unclear: tools are often evaluated on curated examples, formal-verification claims cover only selected components, and little is known about adoption in practice.

This paper studies the state of ZKP security tooling and formal verification across five research questions. We combine a systematization of existing tools, an evaluation of six tools on 70 real-world vulnerabilities, a systematic analysis of formal-verification efforts, and a survey of 48 ZKP practitioners. This methodology lets us assess which languages tools cover, what they detect, which parts of the stack are or could be formally verified, and how practitioners use them. Below, we state the research questions and summarize representative findings.

\par\smallskip\noindent\textbf{\hypertarget{rq:1}{RQ1}: What is the current landscape of ZKP security tools?}
Which DSLs do existing tools support?
What analysis techniques do they use?
Which vulnerability classes can they detect, and which classes remain unsupported?

\emph{Key findings.} Current tools focus almost exclusively on Circom circuits and concentrate on nondeterminism and related underconstraint bugs, while leaving newer DSLs, zkVMs, and other vulnerabilities weakly supported (\autoref{sec:qualitative-tools}).

\par\smallskip\noindent\textbf{\hypertarget{rq:2}{RQ2}: How effective are automated ZKP security tools on real-world vulnerabilities?}
What affects their effectiveness?

\emph{Key findings.} Across 70 real-world bugs, at least one tool detects 45.7\% on isolated circuits, but only 19.6\% on whole-project codebases, with errors, timeouts, and compatibility failures limiting practical effectiveness (\autoref{sec:rq2-tool-effectiveness}).


\par\smallskip\noindent\textbf{\hypertarget{rq:3}{RQ3}: What parts of the ZKP stack are covered by formal verification efforts?} Which systems and properties have been verified? How are models built, proofs discharged, and trusted assumptions defined? What verification gaps remain?

\emph{Key findings.} zkVM verification is converging on Lean and the Sail RISC-V specification, but results remain isolated: they mostly prove constraint soundness, leave key artifacts unverified, and rely on trusted extraction and modeling. Progress requires end-to-end, CI-friendly proofs with precise claims (\autoref{sec:fv-zkvms}).

\par\smallskip\noindent\textbf{\hypertarget{rq:4}{RQ4}: How are ZKP security tools adopted in development and audit workflows?}
Which tools and techniques are used by practitioners?
How are tools integrated into development and audit workflows?
What barriers limit tool adoption?

\emph{Key findings.} Workflows remain human-led even though interactive LLMs are already mainstream, used by 85\% of developers and 83\% of auditors; advanced tools are useful but costly to integrate, often requiring project-specific specifications and manual validation (\autoref{sec:practitioners-perspective}).

\par\smallskip\noindent\textbf{\hypertarget{rq:5}{RQ5}: What gaps do practitioners perceive in security tooling?}
Which vulnerability classes are hardest to detect during development and audits?
Which classes are adequately covered by existing tools, and which remain unsupported?
Which tool characteristics do practitioners prioritize?
How do practitioners view emerging AI/LLM-based assistance?

\emph{Key findings.} Practitioners view underconstrained bugs as the most mature target for current tooling, but identify major gaps around semantic errors, Fiat--Shamir issues, and integrations. They prioritize formal verification, LLMs, and tools with clear reports and guarantees (\autoref{sec:practitioners-perspective}).

\textbf{Availability.}
To support reproducibility and further research, all of our artifacts are publicly available at \url{https://github.com/t-sorger/zkp-security-tools}, including the extended bug dataset, the tool selection process, the tool-evaluation harness and its results, and the anonymized raw survey data. 

\section{Background}
\label{sec:background}

\begin{figure}[t]
    \centering
    \setlength{\tabcolsep}{4mm}%
    \captionsetup[subfigure]{justification=centering, singlelinecheck=false, margin={0pt,50pt}}%
    \begin{tabular}{@{}c !{\color{black!25}\vrule width 0.5pt} c@{}}
        \begin{subfigure}[t]{0.20\textwidth}
            \centering
            \caption{ZK-DSL}
            \label{fig:pipeline-dsl}
            \vspace{1mm}
            \includegraphics[width=\linewidth]{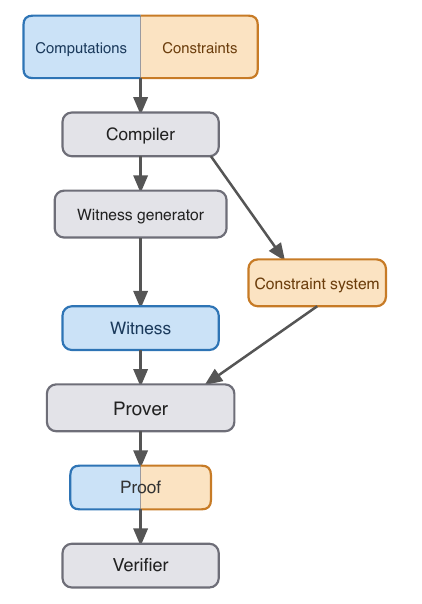}
        \end{subfigure}
        &
        \begin{subfigure}[t]{0.20\textwidth}
            \centering
            \caption{zkVM}
            \label{fig:pipeline-zkvm}
            \vspace{1mm}
            \includegraphics[width=\linewidth]{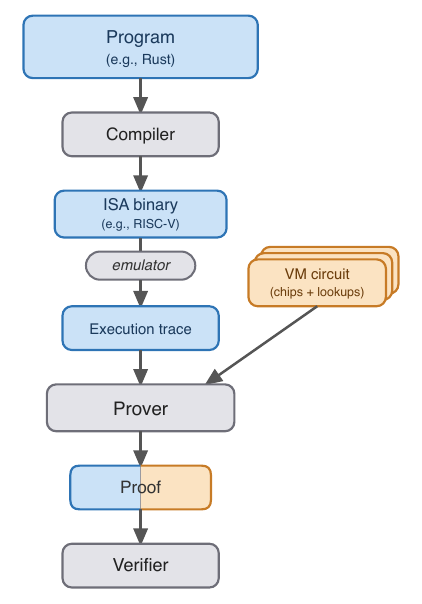}
        \end{subfigure}
    \end{tabular}

    \vspace{1.5mm}
    \includegraphics[width=0.4\textwidth]{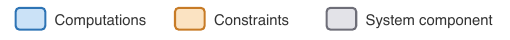}

    \caption{Two ways to develop a ZKP application. A ZK-DSL~(\subref{fig:pipeline-dsl}) compiles a hand-written circuit (computations \emph{and} constraints) into a witness generator and a constraint system. A zkVM~(\subref{fig:pipeline-zkvm}) instead runs an ordinary program on a fixed ISA, checking the resulting execution trace against a fixed circuit.}
    \label{fig:pipelines}
\end{figure}

\point{Zero-knowledge proofs (ZKPs)}
A \textit{zero-knowledge proof} (ZKP) lets a \textit{prover} convince a \textit{verifier} that a statement is true without revealing anything beyond its validity~\cite{goldwasser1985knowledge}. (In practice, proofs may only be succinct rather than zero-knowledge, but the term is still used.)
Typically, this statement asserts that a \textit{computation} was performed correctly on public and optionally private inputs.
To make this provable, the computation is expressed as a set of \textit{constraints}, which are polynomial equations over a finite field.
This is necessary because the proof system works on such equations, not arbitrary programs: any non-polynomial part of the computation, such as a comparison, must be re-encoded as equivalent constraints.
A satisfying assignment of these equations, called a \textit{witness}, then corresponds to a correct run of the computation.
The prover produces a proof that it knows such a witness, and the verifier checks the proof.
Developers build such ZKP programs in two ways, depicted in \autoref{fig:pipelines}.

\point{ZK-DSLs}
The first approach is to write an \textit{arithmetic circuit} by leveraging a domain-specific language (DSL) such as Circom~\cite{belles2023circom}. 
When using such a DSL, the developer writes both the computations and the constraints themselves and must keep them in sync~\cite{chaliasos2024sok, kolozyan2025dcm}.
A compiler then lowers the circuit into a constraint system (e.g., R1CS~\cite{groth2016pairing}), together with a \textit{witness generator} that computes all intermediate values for a given input. Then, the prover combines this witness with the constraints into a cryptographic proof that the verifier checks.

\point{Vulnerabilities}
Developing using ZK-DSLs is error-prone, because a mismatch between the computations and the constraints can break two core properties.
If the constraints are too ``weak'', they accept witnesses that the computation would never produce. This breaks \textit{soundness}, letting a malicious prover convince the verifier of an incorrect statement.
If instead the constraints are too strict, they reject witnesses from honest runs, which breaks \textit{completeness}.
A soundness failure is the most dangerous, as it can let an attacker forge proofs of false statements, leading to issues such as inflation bugs~\cite{wilcox2026orchard}.

\point{zkVMs}
Beyond being a source of bugs, writing constraints ``by hand'' does not scale to large programs and is impractical to maintain as programs change.
A zkVM removes this burden: the developer writes an ordinary program in a conventional language (e.g., Rust), compiles it to a fixed \textit{instruction set architecture} (ISA), and the zkVM proves the program (which is committed to) was executed correctly~\cite{bensasson2014succinct}.
In this model, the circuit no longer encodes the application but the \textit{fetch-decode-execute} loop of a CPU, which proves the execution of \textit{any} program for that ISA.
In a zkVM, the witness generator corresponds to an ISA \textit{emulator}. It first runs the program and records its \textit{execution trace}, which is a step-by-step log of every instruction executed.
This trace is then checked against the constraints, which are split by instruction type into \textit{chips} (e.g., one chip for addition, one for memory).
Beyond these core chips, zkVMs often rely on \textit{precompiles}: dedicated circuits for expensive, frequently used operations (e.g., hashing) that would be prohibitively slow to prove~\cite{risc02026precompiles}.
As each chip is constrained on its own, the proof must additionally tie them
together so they describe a single, consistent execution. \textit{Lookup} and
\textit{permutation} arguments provide this link: a lookup proves that a value
lies in a given set (such as the rows another chip exposes),
while a permutation argument proves that two sequences of values are
rearrangements of one another. Together, they keep a shared state such as memory
coherent across the trace, so that a value read always equals the one last
written to that address~\cite{yang2026zkvm,arun2023jolt}.
Finally, a prover backend turns the satisfied constraints into a proof, which a verifier checks.
This generality comes at a cost: each stage of the zkVM pipeline is a separate trust assumption, and therefore, a separate target of formal verification.

\section{Methodology}

\subsection{Qualitative Methodology} \label{sec:qualitative-methodology}

To dissect the current state of ZKP security tooling
(\rqref{1}/\autoref{sec:qualitative-tools}) and the formal verification
efforts for ZKPs (\rqref{3}/\autoref{sec:fv-zkvms}), we conducted a
qualitative analysis combining systematic review with
iterative expert analysis. 

\point{Corpus construction}
For \rqref{1}, we collected ZKP security analysis tools, i.e., tools that analyze
ZKP circuits for security-relevant bugs. We searched academic publications, GitHub repositories, and blog posts/project documentation. 
For \rqref{3}, we collected formal verification efforts targeting ZKP systems, with a focus on zkVMs as they have been the primary target of FV due to their significance and complexity~\cite{yang2026zkvm}.
We included closed-source tools or verification efforts only when public material provided sufficient detail to classify their targets, techniques, and scopes.
The complete process, initial corpora, and filtering are documented in our artifact.

\point{Classification and validation}
Two authors independently classified each artifact through an 
iterative process.
For \rqref{1}, we classified tools by analysis
target, technique, supported vulnerability classes, source
availability, and automation level. Vulnerability coverage was mapped
to existing ZKP bug taxonomies~\cite{chaliasos2024sok,kolozyan2025dcm}, and support
was marked as full, partial, or unsupported based on papers,
documentation, and source code. For \rqref{3}, we classified formal
verification efforts by verification goal, model-construction method,
proving technique, verification scope, and trusted computing base.
Disagreements were resolved through discussion, and unresolved cases
were reviewed by a third author. 
The corpus and classifications were
revisited as we went over the full corpus.

\subsection{Quantitative Methodology}
\label{sec:quantitative-methodology}

To answer \rqref{2}, we evaluate whether automated ZKP security tools can detect real-world vulnerabilities collected from audits and public disclosures, rather than benchmarks tailored to individual tools. We focus on true-positive detection, i.e., whether a tool identifies a vulnerability. We do not measure false positives, as our goal is to assess current bug-finding capability. We leave a false-positive analysis to future work.

\point{DSL \& Tool Selection}
We focus on Circom as it is not only the DSL that most automated ZKP security tools target~\cite{chaliasos2024sok}, but also one of the most mature and widely deployed DSLs. We selected \emph{all} open-source tools that run non-interactively. For tools with optional specification-driven modes, we use only modes that do not require manual annotations.

\point{Dataset}
We extend the dataset introduced by Chaliasos et al.~\cite{chaliasos2024sok}, which contains real-world ZKP circuit vulnerabilities. 
For each bug, the dataset provides the circuit containing the vulnerable code, metadata about the vulnerable location, and, when available, the original vulnerable project commit.
This lets us evaluate each tool in two settings: in a minimal wrapper that isolates the vulnerable code (\textit{isolated} mode) and in the original project codebase (\textit{project} mode).

We extend the Circom dataset from 29 to 70 bugs by adding high-severity vulnerabilities from audit reports and public disclosures.
All 70 bugs compile in \textit{isolated} mode. 
In \textit{project} mode, 56 bugs compile successfully, while the remaining projects fail because of stale code or deprecated dependencies.

\point{Execution and Metrics}
We developed a harness to process bugs, execute
tools, and handle timeouts. For each run, it stores the tool's raw
output and assigns one of four labels: detected, missed, error, or
timeout. Detected and missed are decided by comparing the tool's
findings against the ground truth.
When a tool reports a vulnerability but does not directly identify the
ground-truth location, two authors manually inspect the raw output and
compare it with the dataset annotation before assigning the final label.
Experiments were executed with a 600-second timeout on a virtual machine with 8 vCPUs, 32\,GB RAM, and an AMD EPYC-Milan CPU running at 2.4\,GHz.

\subsection{Survey Methodology}\label{sec:survey-methodology}
\point{Survey Design}
To answer \rqref{4} and \rqref{5}, we designed a practitioner survey following standard guidelines for survey research~\cite{kitchenham2008survey}, similar to prior software engineering and smart contract security studies~\cite{christakis2016developers,chaliasos2024smart}.
The survey covered security tool usage, perceived tooling gaps, formal verification workflows, and LLM-assisted security practices.

We created two questionnaire variants: one for developers and researchers building ZKP systems, and one for security practitioners auditing them. Both variants were anonymous, all questions were optional, and free-text fields were included where responses could benefit from additional context.

To refine the questionnaires, two authors first drafted the survey independently and then converged on common versions. We then ran two pilot rounds with $N=3$ participants per questionnaire in each round. Pilot feedback was used to clarify wording and adjust multiple-choice options.

\point{Respondent Selection and Demographics}
We targeted practitioners with experience in high-impact ZKP systems, focusing on deployed protocols with substantial usage. To reach this specialized population, we primarily contacted practitioners at research, development, and security organizations through personal channels (e.g., email and instant messaging).

In total, we contacted 200 practitioners. Since many practitioners work across multiple roles, we asked participants to complete the questionnaire most relevant to their work. We received 48 responses, corresponding to a 24\% response rate. \autoref{tab:survey-demographics} summarizes participants' experience.

\begin{table}[t]
\centering
\caption{
    Survey participants' experience.
}
\label{tab:survey-demographics}
\footnotesize
\setlength{\tabcolsep}{3pt}
\renewcommand{\arraystretch}{0.97}
\begin{tabular}{@{}p{0.52\columnwidth}cc@{}}
\toprule
 & \textbf{Developers} & \textbf{Auditors} \\
\midrule
\multicolumn{3}{l}{\textit{Years of ZKP experience}} \\
    \quad Less than 1 & 2 (5.6\%) & 2 (16.7\%) \\
    \quad 1--2 & 10 (27.8\%) & 6 (50.0\%) \\
    \quad 3--5 & 18 (50.0\%) & 4 (33.3\%) \\
    \quad 6+ & 6 (16.7\%) & 0 (0.0\%) \\
\multicolumn{3}{l}{\textit{Types of ZKP systems used (multiple allowed)}} \\
    \quad Circuit DSLs & 29 (80.6\%) & 11 (91.7\%) \\
    \quad zkVM implementations & 23 (63.9\%) & 8 (66.7\%) \\
    \quad zkVM guest programs & 10 (27.8\%) & 2 (16.7\%) \\
    \quad Other / specialized & 3 (8.3\%) & 1 (8.3\%) \\
\midrule
\textbf{Total} & \textbf{36} & \textbf{12} \\
\bottomrule
\end{tabular}
\end{table}

\point{Data Analysis}
We analyzed responses based on question type. For multiple-choice and Likert-scale questions, we report response distributions and percentages over the participants who answered each question. For open-ended questions, two authors reviewed the responses qualitatively and grouped recurring themes through inductive coding. 
We report results separately for developers/researchers and security practitioners where the distinction is relevant.

\section{ZKP Security Analysis Tool Landscape} \label{sec:qualitative-tools}
This section answers \rqref{1} by classifying ZKP security analysis tools by their target, technique, coverage, and usability. 

\begin{table*}[tb]
\centering
\begin{minipage}[t]{0.49\textwidth}
\centering
\caption{Qualitative comparison of ZKP circuit tooling. C = Circom, R = R1CS, CN = Circom/Noir, H = halo2, CH = Circom/halo2. SA = static analysis, DA = dynamic analysis, FV = formal verification, SF = SA+FV, $\sim$ partial support.}
\label{tab:tool-qualitative}
\footnotesize
\setlength{\tabcolsep}{0.6pt}
\renewcommand{\arraystretch}{0.92}
\newcommand{\qtool}[1]{\makebox[0pt][l]{\rotatebox[origin=lb]{72}{\strut #1}}}
\begin{tabular}{@{}p{0.22\linewidth}*{12}{>{\centering\arraybackslash}p{0.058\linewidth}}@{}}
\toprule
\raisebox{2.3em}{\bf Dimension} & \rule{0pt}{4.9em}
\qtool{Circomspect~\cite{trailofbits2024circomspect}} &
\qtool{CCC-Check~\cite{kolozyan2025dcm}} &
\qtool{Pilspector~\cite{Schaeffer2023Pilspector}} &
\qtool{Korrekt~\cite{soureshjani2023halo2}} &
\qtool{zkFuzz~\cite{takahashi2025zkfuzz}} &
\qtool{Zequal~\cite{stephens2025automated}} &
\qtool{ConsCS~\cite{jiang2025conscs}} &
\qtool{Picus~\cite{pailoor2023automated}} &
\qtool{Civer~\cite{isabel2024civer}} &
\qtool{Ecne~\cite{wang2022ecne}} &
\qtool{ZKAP~\cite{wen2024practical}} &
\qtool{AC4~\cite{yang2026ac4}} \\[1pt]
\midrule
Target &
C & IR & PIL & H & CN & C & R & R &
C & R & C & CH \\
Technique &
SA & SA & FV & SF & DA & SF & SF & SF &
FV & FV & SA & FV \\
\midrule
\makecell[l]{Underconstrained} &
\checkmark & \xmark & \checkmark & \checkmark & \checkmark & \checkmark & \checkmark & \checkmark &
\checkmark & \checkmark & \checkmark & \checkmark \\
Overconstrained &
\xmark & \xmark & \xmark & \checkmark & \checkmark & \xmark & \xmark & \xmark &
\xmark & \xmark & \xmark & \checkmark \\
\midrule
Open source &
\checkmark & \checkmark & \checkmark & \checkmark & \checkmark & \checkmark & \checkmark & \checkmark &
\checkmark & \checkmark & \checkmark & \xmark \\
Automatic &
\checkmark & $\sim$ & $\sim$ & \checkmark & \checkmark & \checkmark & \checkmark & \checkmark &
$\sim$ & \checkmark & \checkmark & \checkmark \\
\bottomrule
\end{tabular}
\let\qtool\relax
\renewcommand{\arraystretch}{1.0}

\end{minipage}\hfill
\begin{minipage}[t]{0.49\textwidth}
\centering
\caption{Detection capabilities of tools with respect to vulnerability categories. A single bug can fall under several of them. $\sim$ = partial support.}
\label{tab:tool-comparison}
\footnotesize
\setlength{\tabcolsep}{0.6pt}
\renewcommand{\arraystretch}{0.92}
\newcommand{\ctool}[1]{\makebox[0pt][l]{\rotatebox[origin=lb]{72}{\strut #1}}}
\begin{tabular}{@{}>{\centering\arraybackslash}p{0.055\linewidth}p{0.22\linewidth}*{12}{>{\centering\arraybackslash}p{0.050\linewidth}}@{}}
\toprule
\multicolumn{2}{@{}l}{\raisebox{2.3em}{\bf Vulnerability Class}} & \rule{0pt}{4.9em}
\ctool{Circomspect~\cite{trailofbits2024circomspect}} &
\ctool{CCC-Check~\cite{kolozyan2025dcm}} &
\ctool{Pilspector~\cite{Schaeffer2023Pilspector}} &
\ctool{Korrekt~\cite{soureshjani2023halo2}} &
\ctool{zkFuzz~\cite{takahashi2025zkfuzz}} &
\ctool{Zequal~\cite{stephens2025automated}} &
\ctool{ConsCS~\cite{jiang2025conscs}} &
\ctool{Picus~\cite{pailoor2023automated}} &
\ctool{Civer~\cite{isabel2024civer}} &
\ctool{Ecne~\cite{wang2022ecne}} &
\ctool{ZKAP~\cite{wen2024practical}} &
\ctool{AC4~\cite{yang2026ac4}} \\
\midrule
\multirow{6}{*}{\rotatebox[origin=c]{90}{\bf Underconstrained}} & \clg{\makecell[l]{Nondeterminism}} &
\clg{$\sim$} & \clg{\xmark} & \clg{\checkmark} & \clg{\checkmark} & \clg{\checkmark} & \clg{\checkmark} & \clg{\checkmark} & \clg{\checkmark} &
\clg{\checkmark} & \clg{\checkmark} & \clg{\checkmark} & \clg{\checkmark} \\
& \makecell[l]{Missing range\\checks} &
$\sim$ & \checkmark & \xmark & $\sim$ & $\sim$ & \xmark & \xmark & $\sim$ &
\checkmark & \xmark & $\sim$ & \xmark \\
& \clg{Division by zero} &
\clg{$\sim$} & \clg{\checkmark} & \clg{\xmark} & \clg{\xmark} & \clg{\checkmark} & \clg{$\sim$} & \clg{$\sim$} & \clg{$\sim$} &
\clg{\xmark} & \clg{\xmark} & \clg{$\sim$} & \clg{\xmark} \\
& \makecell[l]{Circuit logic\\assumptions} &
\xmark & \checkmark & \xmark & \xmark & \xmark & \xmark & \xmark & \xmark &
\checkmark & \xmark & \xmark & \xmark \\
& \clg{Other mismatches} &
\clg{\xmark} & \clg{\checkmark} & \clg{\xmark} & \clg{\xmark} & \clg{$\sim$} & \clg{\checkmark} & \clg{\xmark} & \clg{\xmark} &
\clg{\checkmark} & \clg{\xmark} & \clg{$\sim$} & \clg{\xmark} \\
\midrule
\multicolumn{2}{@{}l}{\bf Overconstrained} &
\xmark & \xmark & \xmark & \checkmark & \checkmark & \xmark & \xmark & \xmark &
\xmark & \xmark & \xmark & \xmark \\
\bottomrule
\end{tabular}
\let\ctool\relax
\renewcommand{\arraystretch}{1.0}

\end{minipage}
\end{table*}

\point{Comparison Dimensions}
\autoref{tab:tool-qualitative} summarizes selected tools along three dimensions. First, it captures the artifact being analyzed (e.g., source-level DSL code, or R1CS) and the analysis technique. Second, it outlines the vulnerability classes the tools can target. Third, it reports which tools are open-source and whether they run automatically without specifications.

\point{Targets and Techniques}
Most tools target Circom, operating at different abstraction levels that correspond to different stages of the DSL pipeline in \autoref{fig:pipeline-dsl}.
Source-level tools, such as \circomspect, analyze Circom code directly, while other tools like \picus operate on constraint systems (R1CS).
The latter is closer to being language-agnostic, but loses the witness-computation logic needed to detect computation-constraint mismatches. To achieve language-agnostic analysis while preserving access to computations, recent work instead targets high-level intermediate representations~\cite{ozdemir2022circ, veridise2026llzk, kolozyan2025dcm}.

The tools leverage three main techniques. Static analysis reasons about code structure and data flow using methods such as pattern matching or taint analysis. It is fast, but often imprecise. Formal verification tools such as \civer prove properties of the constraint system using deduction rules and SMT solving. They provide stronger guarantees when they succeed, but struggle with large circuits and non-linear finite-field arithmetic~\cite{hader2023smt,ozdemir2024grobner,hader2024smtlib,ozdemir2023satisfiability}. To mitigate this, several tools combine static analysis with formal verification~\cite{pailoor2023automated,stephens2025automated}. 
Finally, \zkfuzz applies fuzzing to detect defects~\cite{takahashi2025zkfuzz}.

\point{Vulnerability Coverage}
\autoref{tab:tool-comparison} summarizes tool support for vulnerability classes from prior ZK bug taxonomies~\cite{chaliasos2024sok,kolozyan2025dcm}.
Most tools focus on \textit{nondeterminism}, where constraints allow multiple outputs for the same input~\cite{pailoor2023automated}. This is the most common form of underconstrainedness, but it is not the only one. A circuit is also underconstrained whenever \textit{semantic mismatches} occur: constraints accept witnesses
that do not correspond to valid executions, even when deterministic.

Some tools already extend coverage beyond nondeterminism, but mostly by requiring specifications.
For instance, \civer and \ccccheck detect missing range checks through user-defined preconditions, while \circomspect and \zkfuzz rely on hard-coded checks that do not generalize to arbitrary circuits. 
Complex logic bugs and cryptographic issues remain largely unsupported and are therefore omitted from the table.
  
\point{Practical Usability}
All tools except \acfour are publicly available, but that does not imply push-button use.
While most tools can run automatically, some require user-provided preconditions and postconditions. This specification burden can result in stronger
guarantees and support for a wider range of vulnerabilities, but limits automated use.

\begin{takeawaybox}
\textbf{Takeaways for \rqref{1}}
\begin{itemize}[leftmargin=1.1em, labelsep=0.35em, itemsep=1pt, topsep=2pt, parsep=0pt]
    \item Current ZKP security analysis tools focus almost exclusively on Circom circuits, with limited support for language-agnostic workflows.
    \item Tools mainly target nondeterminism and have weaker coverage for other important bugs, including semantic mismatches and overconstrainedness.
    \item Stronger analyses often require specifications, limiting their use as fully automated tools.
\end{itemize}
\textbf{Call to action:} Security tools should provide broader language support, either through language-agnostic analyses or support for emerging DSLs, and should expand beyond nondeterminism, covering a wider range of vulnerabilities.
\end{takeawaybox}

\section{Tools' Effectiveness} 
\label{sec:rq2-tool-effectiveness}

This section answers \rqref{2} by evaluating existing automated and open-source ZKP security tools against real-world vulnerabilities. 
We evaluate six Circom tools in two modes: \textit{isolated} and \textit{project} mode (see~\autoref{sec:quantitative-methodology} for the details).


\begin{figure}[t]
    \centering
    \includegraphics[width=0.9\columnwidth]{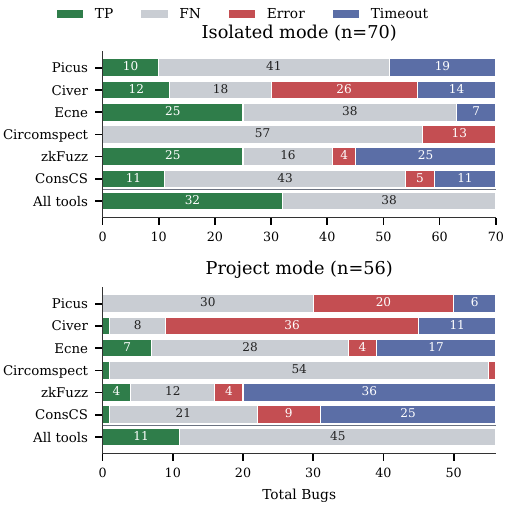}
    \caption{Per-tool outcomes in \textit{Isolated} and \textit{Project} modes.}
    \label{fig:rq2-tool-outcomes}
\end{figure}

\point{Overall effectiveness}
In \textit{isolated} mode, at least one tool detects 32 of the 70 bugs (45.7\%).
Across all executions, the tools produce 83 true positives (TPs). 
However, this hides substantial overlap, as 38 bugs are missed by every tool, while 20 bugs are detected by at least two tools. 
In \textit{project} mode, effectiveness drops sharply: 
at least one tool detects only 11 of the 56 bugs (19.6\%), 
with 14 TPs in total (see~\autoref{fig:rq2-tool-outcomes}).

\point{Detection by root cause}
Detection is concentrated in a narrow subset of bug patterns.
\autoref{tab:rq2-root-cause-detection} groups bugs by their root cause~\cite{chaliasos2024sok}, and counts a bug as detected if any tool reports a TP. 
In \textit{isolated} mode, all \textit{assigned-but-unconstrained} bugs are detected, and tools detect roughly half of the \textit{missing-input-constraint} and \textit{logic-to-constraint-translation} bugs. 
In contrast, no \textit{circuit-design} issues are detected, 
and only one of three \textit{specification misimplementations} is detected. 
The \textit{project} mode results are weaker across nearly every class. 
Importantly, \textit{assigned-but-unconstrained} bugs fall from 100\% to 20\% and \textit{missing-input-constraint} bugs from 45\% to 16\%.

\begin{table}[t]
\centering
\caption{Cumulative detection by ground-truth root cause.}
\label{tab:rq2-root-cause-detection}
\footnotesize
\setlength{\tabcolsep}{3pt}
\begin{tabular}{@{}p{0.50\columnwidth}cc@{}}
\toprule
\textbf{Root cause} &
\textbf{Isolated} &
\textbf{Project} \\
\midrule
\rowcolor{gainsboro}Missing input constraints & 10/22 (45.5\%) & 3/18 (16.7\%) \\
Logic-to-constraint translation & 9/18 (50.0\%) & 3/13 (23.1\%) \\
\rowcolor{gainsboro}Unsafe circuit reuse & 4/10 (40.0\%) & 2/7 (28.6\%) \\
Assigned but unconstrained & 6/6 (100.0\%) & 2/5 (40.0\%) \\
\rowcolor{gainsboro}Circuit design issue & 0/6 (0.0\%) & 0/6 (0.0\%) \\
Arithmetic field issues & 2/4 (50.0\%) & 1/4 (25.0\%) \\
\rowcolor{gainsboro}Spec. misimplementation & 1/3 (33.3\%) & 0/2 (0.0\%) \\
Other programming errors & 0/1 (0.0\%) & 0/1 (0.0\%) \\
\bottomrule
\end{tabular}
\end{table}


\point{Isolated versus project analysis}
The \textit{isolated} mode benchmark is valuable because it standardizes experiments and isolates the vulnerable code, but it can overestimate practical effectiveness. Among the 56 bugs that compile in both modes, 31 are missed in both modes, 6 are detected in both, and 14 are detected only in \textit{isolated} mode.
The common case, therefore, is that large codebases often introduce scalability issues.

\point{Operational limitations}
Tools lack maturity, as they often produce errors. 
Further, as most tools rely on SMT solvers, timeouts occur quite often. 
Specifically, \civer detects 20 TPs in \textit{isolated} mode, but also fails with 26 errors and 14 timeouts. 
In \textit{project} mode, timeouts and errors become more common (see~\autoref{fig:rq2-tool-outcomes}), with most timeouts stemming from the SMT queries the tools perform. A recent line of dedicated finite-field SMT solvers aims to make exactly these queries faster~\cite{ozdemir2023satisfiability,hader2023smt,ozdemir2024grobner,hader2024yices,isabel2026orchestral}. A complementary direction discharges finite-field obligations in a proof assistant instead, sidestepping the theory combination that stalls SMT~\cite{pertseva2026bitmodeq}.

Lightweight tools are more robust but less effective. For instance, \circomspect never times out, yet it detects only one TP.
Although we do not systematically measure false-positive rates in this benchmark, our use of the tools on real codebases revealed that, as expected, \ecne and \circomspect are substantially noisier than the remaining tools, often producing many false positives.

\begin{takeawaybox}
\textbf{Takeaways for \rqref{2}}
\begin{itemize}[leftmargin=1.1em, labelsep=0.35em, itemsep=1pt, topsep=2pt, parsep=0pt]
    \item ZKP security tools have practical value, as they detect critical vulnerabilities, so developers should use them.
    \item The tools are not yet mature. Errors and timeouts are common, while effectiveness drops when analyzing large codebases, suggesting that tools are currently better suited to focused code-fragment analysis.
    \item Results differ substantially from tool-specific benchmarks presented in their research papers, highlighting the bias introduced by narrow evaluation suites.
\end{itemize}
\textbf{Call to action:} ZKP security tools need better engineering, broader vulnerability coverage, and evaluation on real-world datasets. Improvements in SMT solving and complementary analyses could substantially reduce timeouts and make more sophisticated analyses practical.
\end{takeawaybox}

\section{Formal Verification of ZKP Systems} \label{sec:fv-zkvms}
While the tools in \autoref{sec:qualitative-tools} detect bugs in circuits,
a complementary line of work aims to prove their absence through formal verification (FV).
This is particularly relevant for general-purpose ZKP systems such as zkVMs~\cite{yang2026zkvm}, which we focus on here as they also subsume circuit verification.
A zkVM is far more complex than a single circuit, and it concentrates risk: a single zkVM can be a point of failure for many independent applications. For the same reason, however, verifying one zkVM can secure a whole ecosystem.

Automated tools (e.g., \picus) can be used to verify specific properties, but do not cover the full stack: reasoning about ISA semantics, soundness and completeness of semantics, the provable security of cryptographic specifications, and the correctness of cryptographic code against those specifications.
This has led to broad adoption of proof assistants, often Lean, although Rocq~\cite{certik2024zkwasm,formalland2026keccak}, ACL2~\cite{kwan2026jolt}, and EasyCrypt~\cite{nethermind2025zksync} have also been used.
Coordinated efforts such as the Ethereum Foundation's verified-zkEVM project~\cite{ethereumfoundation2025verified} pool resources and tooling, while parallel standardization of the zkVM target itself (i.e., a common RISC-V profile)~\cite{ethact2026zkvmstandards} has helped streamline verification efforts.
We therefore center this section on zkVMs, examining which systems have been verified and for which properties, how their formal models are built, which techniques are used, and what they leave in the trusted computing base.
\autoref{fig:zkvm-stack} depicts the zkVM stack and its verification goals. \autoref{tab:fv-tasks} summarizes the verification efforts to date.

\usetikzlibrary{positioning,arrows.meta,fit,backgrounds,calc}
\pgfdeclarelayer{verification}
\pgfsetlayers{background,verification,main}
\definecolor{sndblue}{rgb}{0.10,0.32,0.60}
\definecolor{cmporange}{rgb}{0.80,0.42,0.05}
\definecolor{specblue}{rgb}{0.20,0.40,0.70}
\definecolor{verifyteal}{rgb}{0.00,0.50,0.45}
\definecolor{implbox}{HTML}{FDFBF7}
\definecolor{implband}{HTML}{F6F1E7}
\definecolor{impltext}{HTML}{8A6E3A}
\providecommand{\sbadge}{\textcolor{sndblue}{\textbf{S}}}
\providecommand{\cbadge}{\textcolor{cmporange}{\textbf{C}}}
\providecommand{\legendimpl}{\tikz[baseline=0.1ex]{\draw[black!60, rounded corners=1pt, fill=implbox] (0,0) rectangle (2ex,1.6ex);}}
\providecommand{\legendspec}{\tikz[baseline=0.1ex]{\draw[specblue, dashed, rounded corners=1pt, fill=specblue!6] (0,0) rectangle (2ex,1.6ex);}}
\providecommand{\legendverify}{\tikz[baseline=-0.6ex]{\draw[->, >=Stealth, semithick, verifyteal, densely dashed] (0,0) -- (3ex,0);}}
\providecommand{\legendflow}{\tikz[baseline=-0.6ex]{\draw[->, >=Stealth, semithick, black!60] (0,0) -- (3ex,0);}}
\providecommand{\legendiface}{\tikz[baseline=-0.6ex]{\draw[<->, >=Stealth, semithick, black!50, densely dotted] (0,0) -- (4.8ex,0);}}
\begin{figure*}[t]
\centering
\resizebox{\textwidth}{!}{%
\begin{tikzpicture}[
  yscale=0.76,
  >=Stealth,
  font=\footnotesize,
  impl/.style={draw=black!60, rounded corners=2pt, align=center, fill=implbox,
               inner sep=2.5pt, minimum height=8mm, text width=20mm},
  spec/.style={draw=specblue, dashed, rounded corners=2pt, align=center,
               fill=specblue!6, inner sep=2.5pt, minimum height=8mm, text width=26mm},
  tag/.style={font=\scriptsize, text=black!80, align=center, inner sep=1pt},
  comp/.style={font=\scriptsize\itshape, text=black!70, fill=white, inner sep=1pt, align=center},
  flow/.style={->, semithick, draw=black!60, rounded corners=3pt,
               shorten >=2pt, shorten <=2pt, line cap=round},
  verify/.style={->, semithick, draw=verifyteal, densely dashed,
                 rounded corners=3pt, shorten >=2pt, shorten <=2pt, line cap=round},
  iface/.style={<->, semithick, draw=black!50, dotted, rounded corners=3pt,
                shorten >=2pt, shorten <=2pt, line cap=round},
]
\node[spec] (isaspec) at (5.0,4.7) {ISA spec: RISC-V Sail \\ extracted to Lean};
\node[spec] (prespec) at (8.8,4.7) {Precompile\\function spec};
\node[spec, text width=32mm] (pssp) at (12.6,4.7) {Proof-system spec:\\IOP $\cdot$ PCS $\cdot$ Fiat--Shamir};
\node[spec, text width=22mm] (guestspec) at (2.15,4.7) {Guest program\\spec};
\node[impl] (guest) at (-0.35,0) {Guest\\program};
\node[impl] (isa)   at (2.4,0) {ISA\\program};
\node[impl] (wit)   at (4.0,2.6) {Witness gen.\\(emulator)};
\node[impl] (con)   at (6.0,0) {Constraints};
\node[impl] (pre)   at (8.8,-2.6) {Precompile\\circuits};
\node[impl] (prover)at (12.6,1.3) {Prover\\(CPU / GPU)};
\node[impl] (ver)   at (12.6,-1.3) {Verifier\\(CPU)};
\node[impl] (rec)   at (15.8,0) {Recursion\\(verifier-in-circuit)};
\node[impl] (chain) at (18.6,0) {On-chain\\verifier};
\node[tag, below=0.5mm of guest] {};
\node[tag, below=0.5mm of con]   {CC \,\textperiodcentered\, \sbadge\cbadge};
\node[tag, below=0.5mm of wit, xshift=3mm]   {WC \,\textperiodcentered\, \cbadge};
\node[tag, below=0.5mm of pre]   {CC \,\textperiodcentered\, \sbadge\cbadge};
\node[tag, below=0.5mm of pssp] {PS, PC \,\textperiodcentered\, \sbadge\cbadge};
\node[tag, below=0.5mm of prover]{PC \,\textperiodcentered\, \cbadge};
\node[tag, below=0.5mm of ver]   {VC \,\textperiodcentered\, \sbadge};
\node[tag, below=0.5mm of rec]   {VC \,\textperiodcentered\, \sbadge\cbadge};
\node[tag, below=0.5mm of chain] {VC \,\textperiodcentered\, \sbadge};
\begin{pgfonlayer}{verification}
\draw[verify] (guest.north east) -- (guestspec.south);
\draw[verify] (isa.north west) to[out=105,in=-125] (isaspec.south west);
\draw[verify] (wit.north) -- ([xshift=-10mm]isaspec.south);
\draw[verify] (con.north) -- (con.north |- isaspec.south);
\draw[verify] (pre.north) -- (prespec.south);
\draw[verify] ([xshift=-10mm]prover.north) -- ([xshift=-10mm]pssp.south);
\draw[verify] (ver.north east) to[out=55,in=-90] ([xshift=10mm]pssp.south);
\draw[verify] (rec.north east) to[out=105,in=0] (pssp.east);
\end{pgfonlayer}
\draw[flow] (guest)--(isa);
\draw[flow] (isa)--(wit);
\draw[flow,<->] (wit.south east) to[out=-35,in=125]
    node[comp,pos=0.46, xshift=0.4mm]{consistency} (con.north west);
\node[comp] at (1.0,0.45) {compiler};
\draw[flow] (con.east) to[out=10,in=180] (prover.west);
\draw[flow] (con.east) to[out=-10,in=180] (ver.west);
\draw[flow] (wit.east) to[out=20,in=150] (prover.north west);
\draw[flow] (prover)--(rec);
\draw[flow] (ver)--(rec);
\draw[flow] (rec)--(chain);
\draw[iface] (isa.south) to[bend right=22] node[comp,below,pos=0.5,yshift=1.5mm,xshift=-2mm]{ecall / libc} (pre.west);
\begin{scope}[on background layer]
  \fill[specblue!8, rounded corners=4pt] (-1.5,3.65) rectangle (20.4,5.6);
  \fill[implband, rounded corners=4pt] (-1.5,-3.65) rectangle (20.4,3.2);
\end{scope}
\node[anchor=north west, specblue, font=\bfseries\scriptsize, align=left] at (-1.45,5.55) {SPECIFICATION\\(reference)};
\node[anchor=north west, text=black!70, font=\bfseries\scriptsize, align=left] at (-1.45,3.1) {IMPLEMENTATION\\(what runs)};
\end{tikzpicture}}
\par\smallskip
\begin{tcolorbox}[
  enhanced,
  colback=black!4, colframe=black!40, boxrule=0.6pt, arc=0.8mm,
  left=5pt, right=5pt, top=2pt, bottom=2pt,
  title=Key, fonttitle=\bfseries\footnotesize,
  coltitle=black!80, colbacktitle=black!10,
  fontupper=\footnotesize,
  before skip=4pt, after skip=2pt,
]
\textbf{Boxes:}~\legendimpl~implementation/artifact (runs)\quad \legendspec~specification (reference)\qquad \textbf{Arrows:}~\legendflow~data flow\quad \legendverify~verification link to spec\quad \legendiface~runtime interface\\
\textbf{Goals:}~CC constraint correctness \quad WC witness--constraint consistency\quad PS proof-system soundness\quad VC verifier correctness\quad PC prover completeness\\
\textbf{Properties:}~\sbadge~soundness\quad \cbadge~completeness\quad \sbadge\cbadge~both
\end{tcolorbox}
\caption{The zkVM verification stack: a \textcolor{specblue}{specification} band (reference models) over an \textcolor{impltext}{implementation} band (what runs), linked by compiler and extraction steps. Each artifact is labeled with its verification goal and its relevance to soundness/completeness.}
\label{fig:zkvm-stack}
\end{figure*}
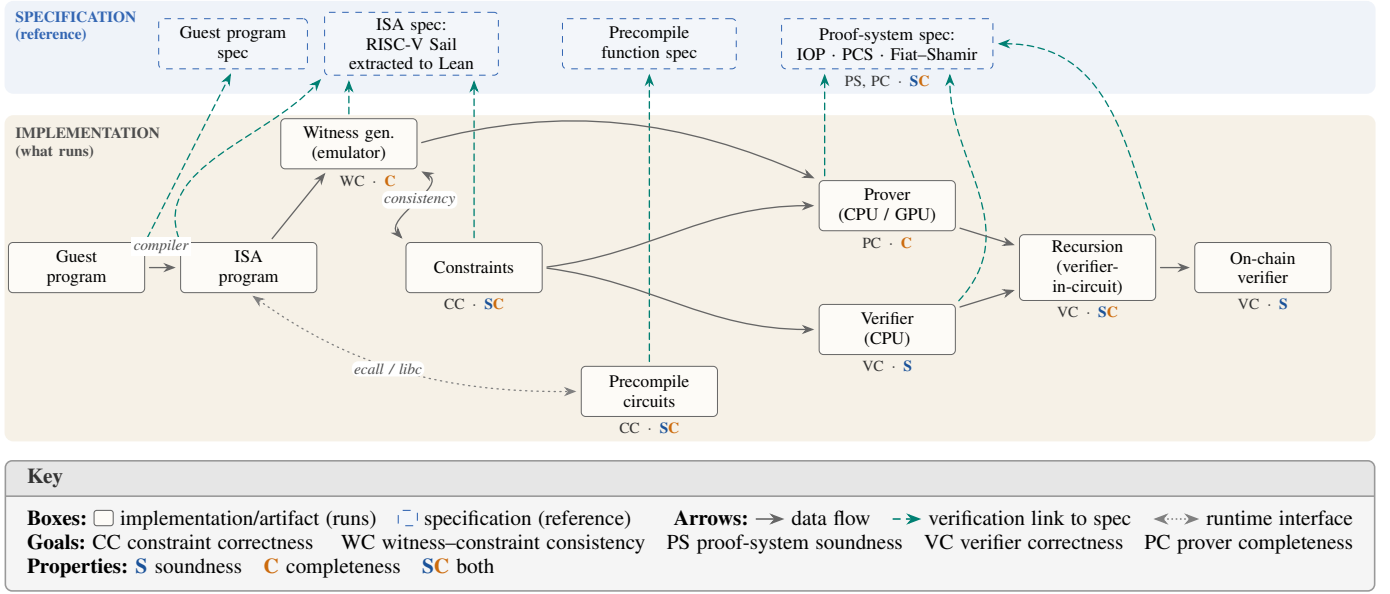

\begin{table*}[t]
    \centering
    \caption{Verification efforts across the zkVM stack up to date. \textit{Verification goal} refers to the goals defined in \autoref{sec:fv-approaches}, with the targeted constraints in brackets for constraint correctness. \textit{Model} lists the model-creation approaches (\autoref{sec:fv-model}). \textit{Proving technique} and \textit{Tooling} report the techniques and tools these efforts employ. ITP = interactive theorem proving.}
    \label{tab:fv-tasks}
    \footnotesize
    \setlength{\tabcolsep}{4pt}
    \begin{tabular}{@{}l >{\raggedright\arraybackslash}p{5.4cm} l >{\raggedright\arraybackslash}p{4.2cm}@{}}
    \toprule
    \bf Verification goal & \bf Model & \bf Proving technique & \bf Tooling \\
    \midrule
    \multicolumn{4}{@{}l}{\textbf{Soundness}}\\
    \addlinespace[1pt]
    \quad Constraint correctness (ISA) & Extraction~\cite{succinct2025sp1,openvm2026fv}, Translation~\cite{veridise2025sp1,veridise2025risc0}, Hand-written~\cite{kwan2026jolt,certik2024zkwasm,avigad2026cairo} & ITP; SMT & Lean, Rocq, ACL2; Picus-AuditHub \\
    \addlinespace[1.5pt]
    \quad Constraint correctness (precompiles) & Extraction~\cite{nethermind2025halva,openvm2026precompiles}, Translation~\cite{formalland2026keccak,veridise2025risc0} & ITP; SMT & Lean, Rocq; Picus-AuditHub \\
    \addlinespace[1.5pt]
    \quad Proof-system soundness & Hand-written~\cite{tuma2026vcvio} & ITP & Lean (ArkLib, VCVio) \\
    \addlinespace[1.5pt]
    \quad Verifier correctness & Hand-written~\cite{nethermind2025zksync} & ITP & EasyCrypt \\
    \midrule
    \multicolumn{4}{@{}l}{\textbf{Completeness}}\\
    \addlinespace[1pt]
    \quad Constraint correctness (ISA) & Correctness by construction~\cite{dellimmagine2025clean} & ITP & Clean \\
    \addlinespace[1.5pt]
    \quad Witness-constraint consistency & Correctness by construction~\cite{dellimmagine2025clean}, Hand-written~\cite{ke2025ziver} & ITP; SMT & Clean; ZIVER \\
    \bottomrule
    \end{tabular}
\end{table*}

\subsection{Verification goals}\label{sec:fv-approaches}

A zkVM composes several artifacts:
a guest program compiled to the ISA, the arithmetization that encodes the ISA as polynomial constraints, precompiles, the cryptographic proof system, and a recursion layer that aggregates proofs for segments of an execution trace into a final one.
Verifying a zkVM means establishing end-to-end soundness and completeness guarantees, with current work typically focusing on soundness.

\point{Soundness}
Soundness means the verifier never accepts proofs of false statements.
It depends on the constraints, the proof system, and the verifier: the witness generator and prover are untrusted.
This reduces the verification scope to the following:
\fvitem{Constraint correctness, soundness direction (CC-S)}
Every witness satisfying the constraints corresponds to an execution of the ISA (constraints $\Rightarrow$ ISA), checked against a reference specification, such as the RISC-V Sail model extracted to a proof assistant. The same obligation applies to each \textit{precompile} against its function specification. This subsumes determinism, which many automated tools target.
\fvitem{Proof-system soundness (PS)}
The cryptographic argument prevents a malicious prover from forging a proof of a false statement that an honest verifier would accept, for some security parameter (e.g., 128 bits).
\fvitem{Verifier correctness (VC)}
The verifier rejects invalid proofs. The verifier used for recursive aggregation carries the same obligation, but is implemented via constraints.
Any use of a zkVM also involves a guest program, which must be compiled correctly or verified at the ISA level so that the compiler drops out of the trusted computing base and the intended statement is the one proven by the zkVM. Given that a zkVM may allow for any guest program to be proven, and that verifying the guest program is a traditional program verification task, we treat this as a mostly separate concern in this section. 

\point{Completeness}
Completeness requires that every honest execution of the guest program results in a proof accepted by an honest verifier.
While soundness is typically the main concern, completeness is essential in some deployments; a completeness failure could, for example, cause liveness issues in a blockchain that settles via validity proofs for blocks.

Unlike soundness, completeness depends on the witness generator and prover actually producing the satisfying witness and a valid proof, which draws in a much larger part of the stack, including compilers. It decomposes into the following.
\fvitem{Constraint correctness, completeness direction (CC-C)}
For every ISA execution, there is a witness that satisfies the constraints (ISA $\Rightarrow$ constraints).
\fvitem{Witness-constraint consistency (WC)}
The witness generator produces a satisfying witness for every execution. The constraints alone can only show that a satisfying witness exists.
\fvitem{Prover completeness (PC)}
The prover produces a proof the verifier accepts. Likewise for the recursion prover, although the proven statement is different.

While constraint DSLs usually derive both constraints and witness computation from one circuit description, zkVMs often optimize the witness generator separately, so WC may cover an artifact untouched by constraint-correctness proofs.

\subsection{Making the system verifiable}\label{sec:fv-model}
To verify any zkVM component, one must have the relevant formal specifications, the ability to precisely state what must be verified, and the ability to model the verification target in a proof assistant.
Specification and modeling of the target largely determine the trusted computing base beyond the verification environment itself, and must therefore be validated and tested. Otherwise, one risks technically correct proofs that say nothing meaningful about the real system.

\point{Model creation}
We distinguish four modeling approaches. The choice is primarily about the gap between the verified model and the code that actually runs, and what is trusted to bridge it. Whether the model is produced by a tool or by hand is secondary, affecting how much code structure survives and how costly the result is to maintain.
\fvitem{Constraint extraction (E)}
In the case of constraints, it is possible to intercept the circuit-builder API (e.g., Plonky3's \texttt{Air} trait) and export the constraints to a proof assistant.
As extraction is automated, the model closely mirrors what the prover actually evaluates, and the main trust assumption is the correctness of the extraction tool.
Nethermind's CertiPlonk~\cite{nethermind2025certiplonk}, OpenVM's constraint-correctness proofs~\cite{openvm2026fv}, and the original SP1 Hypercube proofs~\cite{succinct2025sp1} follow this approach. Extraction is lossy: reading constraints from the builder yields flat polynomials and lookup interactions, losing the code structure that generated them and making proofs longer, harder, and less reusable across similar instructions. Extraction can also bridge approaches: SP1's latest effort extracts constraints but expresses them in Clean~\cite{mitschabaude2026clean}, combining (E) with (CbC).

\fvitem{Translation via a compiler IR (T)}
A compiler lowers the artifact (typically constraints or Rust source code) through a reusable intermediate representation to a verification backend. This preserves more high-level structure than an extractor (e.g., witness-generation logic), but makes the compiler pipeline a major trust assumption that requires extensive validation.
On the constraint side, frameworks such as LLZK~\cite{veridise2026llzk} lower circuits through MLIR dialects to multiple backends. These include an SMT solver (e.g., \picus~\cite{pailoor2023automated}, integrated into RISC Zero's CI/CD pipeline~\cite{veridise2025risc0}) and a proof-assistant backend. The backend is therefore an orthogonal choice while the compiler/IR pipeline joins the trusted base in place of an extractor.
On the Rust side, it can be translated to a proof assistant, whether by \texttt{rocq-of-rust}~\cite{formalland2026keccak} (supplemented by manual work and validated by comparing the Rocq and Rust outputs) or via the Charon frontend~\cite{ho2025charon}.
A limitation of automated Rust-to-proof-assistant tools is that they translate only a fragment of Rust\footnote{For example, see \url{https://github.com/cryspen/hax/labels/unsupported-rust}.} 
and that translation is inherently trusted as there are no complete formal Rust semantics to prove against.
Highly optimized cryptographic code not written with these tools in mind can therefore be hard to extract to any proof assistant: SIMD intrinsics, \texttt{unsafe}, and inline assembly are not supported~\cite{ho2022aeneas, bhargavan2024hax} and are treated as opaque. Hence, only a subset of the code of a prover like Plonky3 reaches the proof assistant.
There are other Rust verification tools, including Verus~\cite{lattuada2023verus} (trialed on zkVM constraints~\cite{certik2026zkvmverus}), which do not extract to a proof assistant and fit self-contained Rust components well. However, they are less suited to certain tasks such as verifying a verifier, which must ideally bridge a Rust implementation to a provably secure proof-system specification that lives in a proof assistant.

\fvitem{Hand-written model (M)}
A human (or agent) writes the model directly in a proof assistant. For example, CertiK manually re-implemented zkWasm's circuit logic in Rocq~\cite{certik2024zkwasm}. This is laborious but flexible. Model-implementation correspondence must be established separately, typically by testing rather than formal equivalence. Jolt's ACL2 model~\cite{kwan2026jolt} did so by exhaustively comparing subtable outputs against Rust for all inputs when feasible (at most $2^{16}$ entries), and by random testing otherwise. StarkWare's Cairo effort similarly built a manual Lean model of VM execution semantics~\cite{avigad2026cairo}, complemented by a proof-producing compiler.

\fvitem{Correctness by construction (CbC)}
Instead of deriving a model from existing code, the circuit is written and proven correct directly in the proof assistant. This removes the model-implementation gap, but requires (re)writing the artifact there. It is only practical when either fast code can be generated from it (as in fiat-crypto~\cite{erbsen2019fiat}) or the artifact is not performance-sensitive, as with constraints. Clean~\cite{mitschabaude2026clean}, implemented in Lean, follows this route and proves both CC and WC.

In short, automated approaches stay closest to the deployed system but concentrate trust in unverified tooling, and extraction may discard code structure. Hand-written models recover structure and cover more of the system, but need separate validation and maintenance. CbC removes the model-code gap, but requires a proof-assistant implementation to be practical from a software-engineering perspective.

\point{Proving techniques}
How a property is discharged also shapes the trusted computing base. Interactive theorem proving is the most expressive option and yields kernel-checked proofs, but proof effort is high (though tactics and AI help) and limited automation hinders CI integration when proofs break. Automated solvers such as SMT require less effort, but have narrower scope (e.g., only target determinism) and can struggle with ZKPs (\autoref{sec:qualitative-tools}). Proof assistants can sidestep this issue: BitModEq, for instance, proves finite-field-to-bitvector equivalences as a Lean tactic, outperforming general-purpose SMT solvers on real zkVM benchmarks~\cite{pertseva2026bitmodeq}.

\point{Trusted computing base}
Every formal verification result holds only relative to a trusted computing base, which has two distinct sources.
The first is the verification target: which parts of the stack were modeled, which reference specification serves as the ground truth, and the gap between the proven model and the code that actually runs. This gap is shaped directly by the model-creation choice (\autoref{sec:fv-model}). Extraction (E) and compiler translation (T) each insert an unverified tool; hand-written models (M) assume that the model, written by a human or agent, matches the implementation; correctness-by-construction (CbC) has no separate model that can diverge but may be less performant. Any processing below the level at which the artifact is verified (e.g., compilation) is also trusted not to alter semantics. Wherever the model is not directly extracted, this gap must be \emph{validated} rather than proven.

The second source is the verification tools themselves, and it is the easier of the two to under-report.
It includes the proof assistant's trusted kernel, together with the axioms and library lemmas invoked, and any unverified solvers or decision procedures trusted to be correct. Solvers are a subtle case: even when they produce a kernel-checkable certificate for the proof assistant, the encoding of the proof obligation passed to them may still be trusted.
It also includes the theorem statements themselves: a proof of a vacuous or misstated theorem passes the kernel while establishing nothing, so what is proven is as much part of the TCB as the machinery that checks it.
Finally, an effort may simply assume a property rather than prove it. OpenVM's constraint-correctness proofs, for example, assume the soundness of the lookup/bus argument that ties its chips together~\cite{openvm2026fv}. 
A subtler cost is that large verification toolchains exist partly just to move artifacts from one representation into a proof assistant or SMT-LIB, never discharging an obligation themselves while still enlarging the TCB and the maintenance burden. This is part of why tighter integration of development tooling with formal verification, and minimizing the set of trusted tools and extractions, is attractive.

The TCB is therefore not a fixed baseline, but a function of both the system and the verification approach, and two verification outputs can rest on very different assumptions. This is a familiar lesson from software verification: even CompCert, the flagship verified C compiler, required an explicit and careful accounting of its trusted base, its specification, its extraction to executable code, and unverified support code~\cite{monniaux2022compcert}.

\subsection{Verification Gaps and Risks}
\label{sec:fv-gaps}

\point{Isolated components}
Constraint soundness is the most mature goal (\autoref{tab:fv-tasks}), with proofs that have covered many zkVMs~\cite{openvm2026fv,kwan2026jolt,certik2024zkwasm,avigad2026cairo}. It is, however, only one direction of one layer of the stack, and significant gaps remain elsewhere.
No zkVM has end-to-end formal verification, and verified components are isolated. Constraint-correctness proofs cover instruction semantics, but the witness generator, the lookup/permutation arguments, the proof-system backend, and the on-chain verifier remain largely unverified.

\point{Shifted complexity}
As the stack is large, verification complexity is often shifted between components rather than removed. Simplifying one component can burden another: lookup arguments are advertised as leading to simpler zkVMs and simpler constraints~\cite{arun2023jolt, thaler2024snark}, but end-to-end verification must then also cover the lookup argument's specification, its underlying cryptography, and the well-formedness of its tables.

Implementation choices can also sidestep whole tasks. Working directly with assembly, say, drops the compiler from the obligations, at the cost of different tools and specifications. In both cases the verified boundary can shrink even as constraint-level proofs become more tractable, with complexity reappearing in the trusted computing base. 

\point{Witness generator}
The witness generator is a core unverified component.
WC could be checked automatically~\cite{stephens2025automated}, but little work applies this to zkVMs, where it remains a research prototype~\cite{ke2025ziver}, and extracting witness generation code to a proof assistant has not yet been done.


\point{Precompiles}
Precompiles also remain largely unverified. Only individual precompiles have been verified in some form, such as Plonky3's Keccak circuit~\cite{formalland2026keccak} and the determinism of RISC Zero's Keccak circuit~\cite{veridise2025risc0}. In one such attempt, Nethermind found a critical bug in Scroll's Halo2 Keccak circuit~\cite{nethermind2025halva}. Beyond these, most precompiles remain unverified across all zkVMs.
A single zkVM may ship precompiles for several cryptographic primitives.
Precompiles can also be generated automatically~\cite{powdr2025autoprecompiles} and synthesized from frequently executed guest code, thereby multiplying their number into an open-ended, shifting set that cannot be chased circuit by circuit. This shifts what kind of verification is required to verify the generator. Assuming one can verify the ISA constraints, the task of verifying autoprecompiles is simply translation validation, which is far more amenable to automation.

\point{Proof system and verifier}
As for the proof system, ArkLib~\cite{verifiedzkevm2025arklib} and VCVio~\cite{tuma2026vcvio} aim to formalize its cryptographic building blocks. These include the sum-check protocol~\cite{lund1992sumcheck}, FRI~\cite{bensasson2018fri}, WHIR~\cite{arnon2025whir}, and Fiat--Shamir~\cite{chiesa2025fiat}. Bridging such a library with production-grade implementations has not yet been done.
The best verified verifier result so far targeted ZKsync Era (an Ethereum zk-rollup), and proved only that the on-chain verifier contract matches a specification for which no soundness properties have been verified~\cite{nethermind2025zksync}. 
Similarly, the lookup arguments and memory consistency checks~\cite{habock2022logup,setty2025twistshout} are almost entirely unverified, but could be specified in a library such as ArkLib.

\point{Maintenance}
There is an economic dimension to all this. Verification can be lengthy and expensive -- it is often done by external contractors -- while zkVMs evolve rapidly, sometimes with multiple releases per year. Hence, several efforts we cite target code that is now deprecated, making verification seem wasteful. The key challenge is therefore maintaining verification status as specifications and implementations change, as in RISC Zero's \picus CI/CD integration~\cite{veridise2025risc0}. AI is making it more feasible to keep pace, but the usual pitfalls remain: whether the specification and model faithfully capture the system, and whether verification is integrated into development. Guarantees not checked on every commit silently decay, while AI-generated proofs remain a poor fit for CI.

\point{Vague claims}
As formal verification becomes more accessible through AI and stands to benefit from greater expertise in the domains to which it is applied, it will be important to remain clear about results. Claims have sometimes been vague, and issues have been found upon close inspection~\cite{gunton2026sp1fv, kobeissi2026theatre}.

\begin{takeawaybox}
\textbf{Takeaways for \rqref{3}}
\begin{itemize}[leftmargin=1.1em, labelsep=0.35em, itemsep=1pt, topsep=2pt, parsep=0pt]
    \item Verification is converging on Lean 4, RISC-V Sail, and libraries such as Clean and ArkLib, but broad adoption remains open.
    \item Verified components remain isolated: constraint soundness is the mature core, while witness generators, lookup/permutation and memory arguments, proof backends, and verifiers are largely unverified.
    \item Every result depends on its trusted computing base, especially the gap between the verified model and running code, so assumptions and trusted tools must be reviewed explicitly.
    \item Adoption depends on economics and integration: zkVMs evolve quickly, proofs must be cheap to re-establish in CI, and AI lowers proof cost without solving specification fidelity.
\end{itemize}
\textbf{Call to action:} Move from isolated constraint-correctness proofs to repeatable, CI-friendly end-to-end guarantees, and state precisely which properties, components, and versions are verified, and under what assumptions.
\end{takeawaybox}

\section{Practitioners' Perspective}\label{sec:practitioners-perspective}

This section answers \rqref{4} and \rqref{5} through targeted surveys of ZKP practitioners. We present the main highlights here, while complete results are in the supplementary material.

\begin{figure*}[t]
    \centering
    \begin{minipage}[t]{0.63\textwidth}
        \vspace{0pt}
        \centering
        \captionsetup[subfigure]{skip=1pt}
        \begin{subfigure}[t]{0.315\linewidth}
            \centering
            \includegraphics[width=\linewidth]{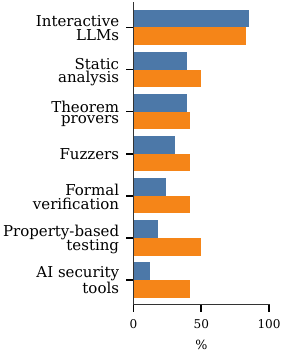}
            \caption{Used tool categories}
            \label{fig:practitioner-used-tools}
        \end{subfigure}\hfill
        \begin{subfigure}[t]{0.30\linewidth}
            \centering
            \includegraphics[width=\linewidth]{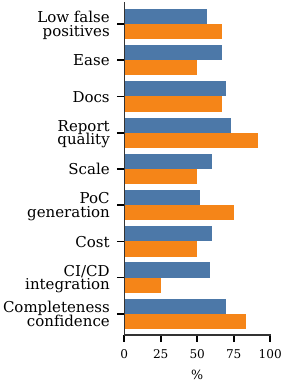}
            \caption{Important tool traits}
            \label{fig:practitioner-tool-traits}
        \end{subfigure}\hfill
        \begin{subfigure}[t]{0.315\linewidth}
            \centering
            \includegraphics[width=\linewidth]{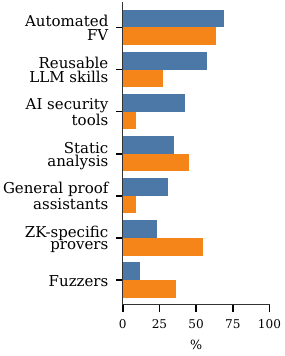}
            \caption{Missing tool categories}
            \label{fig:practitioner-missing-tools}
        \end{subfigure}
        \par\vspace{1pt}
        {\footnotesize
        \textcolor[HTML]{4C78A8}{\rule{0.8em}{0.55em}} Developers\quad
        \textcolor[HTML]{F58518}{\rule{0.8em}{0.55em}} Auditors}
        \caption{Practitioner survey overview.}
        \label{fig:practitioner-overview}
    \end{minipage}\hfill
    \begin{minipage}[t]{0.32\textwidth}
        \vspace{0pt}
        \centering
\captionof{table}{Practitioners' views of vulnerabilities and tools (including LLMs). Hard: hard to detect manually. Covered: adequately covered by tools.}
\label{tab:practitioner-vulnerability-perceptions}
\footnotesize
\setlength{\tabcolsep}{3pt}
\renewcommand{\arraystretch}{1.08}
\begin{tabular}{@{}p{0.52\linewidth}cc@{}}
\toprule
\textbf{Class} & \textbf{Hard} & \textbf{Covered} \\
\midrule
Underconstrained circuits & \cellcolor{cyan!25}72\% & \cellcolor{cyan!24}70\% \\
Arithmetic issues & \cellcolor{cyan!15}44\% & \cellcolor{cyan!26}78\% \\
Semantic mismatches & \cellcolor{cyan!18}53\% & \cellcolor{cyan!10}30\% \\
Proof-system issues & \cellcolor{cyan!21}60\% & \cellcolor{cyan!7}15\% \\
Fiat-Shamir & \cellcolor{cyan!15}44\% & \cellcolor{cyan!8}25\% \\
Crypto primitive misuse & \cellcolor{cyan!11}33\% & \cellcolor{cyan!7}12\% \\
System integration & \cellcolor{cyan!9}26\% & \cellcolor{cyan!7}5\% \\
\bottomrule
\end{tabular}
\renewcommand{\arraystretch}{1.0}

    \end{minipage}
\end{figure*}

\point{Tool adoption and workflows}
As shown in \autoref{fig:practitioner-used-tools}, interactive LLMs are already the most popular across both groups.
Auditors additionally report broader use of specialized tooling,
such as fuzzers, formal verification, and internal AI tools. Beyond which tools they adopt, practitioners also shared their workflows. Manual code review remains dominant, reported by 79\% of developers and 83\% of auditors. While developers rely more on test suites (52\%) and differential testing (41\%), auditors make heavier use of AI tools (67\%) and internal tooling (58\%). The picture is thus not one of push-button security, as tooling is layered around human-led review.

\point{Usefulness and adoption barriers}
When rating usefulness, 57\% of developers and 67\% of auditors rated tools as helpful. 
At the same time, the trait rankings in \autoref{fig:practitioner-tool-traits} show that practitioners weigh how findings are reported as much as what is found.
Report quality and confidence of completeness stand out as the most valued traits across both groups, alongside low false-positive rates and good documentation. Auditors additionally place substantial weight on proof-of-concept generation.
These results indicate that practitioners need tools that not only emit warnings but also clearly report findings. Usefulness, however, does not guarantee adoption: even tools that practitioners rate as helpful remain costly to integrate. High onboarding effort is reported by 57\% of developers and 45\% of auditors. 
Coverage is another barrier: practitioners work across languages and frameworks, which current open-source tools do not support. 
In summary, our survey demonstrates that the dominant adoption barriers are practical, such as integration cost, limited language coverage, and the effort required to write specifications.

\begin{takeawaybox}
    \textbf{Takeaways for \rqref{4}}
    \begin{itemize}[leftmargin=1.1em, labelsep=0.35em, itemsep=1pt, topsep=2pt, parsep=0pt]
        \item Interactive LLMs are already mainstream in ZKP development and auditing, but workflows remain human-led.
        \item Auditors rely more heavily on specialized security tooling than developers.
        \item Practitioners work across many ZKP languages, but most security tools still target only Circom.
        \item Practitioners find tools useful, but advanced tooling still requires high-effort integration, project-specific specifications, and manual validation.
    \end{itemize}
    \textbf{Call to action:} ZKP security tools should not only broaden language support, but also reduce onboarding cost through reusable specifications and integrations that fit both development-time and audit-time use.
\end{takeawaybox}

\point{Perceived vulnerability coverage}
Table~\ref{tab:practitioner-vulnerability-perceptions} reports, for each vulnerability class, the share of practitioners who consider it hard to detect manually and those who consider it adequately covered by current tools.
The results suggest that current tools are perceived as most mature for local circuit-level checks. 
Underconstrained circuits are the most frequently selected hard-to-detect class (72\%), but they are also widely viewed as covered by existing tools (70\%). 
By contrast, proof-system issues combine high perceived difficulty (60\%) with low perceived tool coverage (15\%). Fiat--Shamir bugs, misuse of cryptographic primitives, system integration issues, and semantic errors exhibit similar mismatches. 
The largest gaps thus lie beyond local constraint checks, pointing to specification-centered and FV tools as a promising direction for future work.

\point{Missing tools and unmet needs}
\autoref{fig:practitioner-missing-tools} shows that automated formal verification is the most requested missing category for both groups.
Developers also request reusable LLM skills and AI security tools, while auditors report stronger demand for domain-specific interactive provers and fuzzers. 
A similar gap appears for LLMs: while respondents find them useful for PoC validation and non-ZKP application code, they remain weak precisely for circuit- and proof-system-level vulnerability discovery. 
Even so, many report significant productivity gains, and over 80\% of both groups expect AI auditing to outperform 75\% of security researchers within three years. 
Further, when using formal verification tools, a similar majority (over 80\%) of both groups are at least somewhat concerned
about the trusted computing base. 
Auditors also report that communicating a tool's security guarantees or limitations to clients often requires substantial effort. 
This suggests that future tooling must also improve transparency around assumptions and guarantees.

\begin{takeawaybox}
    \textbf{Takeaways for \rqref{5}}
    \begin{itemize}[leftmargin=1.1em, labelsep=0.35em, itemsep=1pt, topsep=2pt, parsep=0pt]
        \item Current tools are perceived as most mature for local circuit-level bugs, especially underconstrained issues.
        \item Practitioners identify major gaps for semantic mismatches, proof-system unsoundness, Fiat--Shamir bugs, cryptographic primitive misuse, and system integration.
        \item Demand is strongest for automated formal verification, domain-specific provers, reusable LLM skills, and tools with clear reports and auditable guarantees.
    \end{itemize}
    \textbf{Call to action:} The ZKP tooling community should prioritize specification-centered and semantics-aware (AI) tools that make guarantees explicit and practical to validate.
\end{takeawaybox}

\section{Discussion}

\point{From local checks to system assurance}
The four parts of our study converge on a common boundary. 
Automated analyses are strongest on local constraint properties, yet their effectiveness drops when moving from isolated circuits to full projects.
Formal verification offers stronger guarantees, but mostly for the soundness of selected components, while practitioners identify semantic, proof-system, cryptographic, and integration failures as major gaps.
Those observations show that adding further local detectors alone will not close the assurance gap.
Future work should connect layers through obligations such as computation--constraint consistency, interface invariants, translation validation, and component composition, while making uncovered layers explicit.

\point{Specifications as shared security infrastructure}
Moving beyond local properties requires making intended behavior explicit.
This connects three findings: broader circuit analyses require user-provided specifications, formal-verification guarantees depend on the fidelity of theorem statements and models, and practitioners report specification work as an adoption barrier. 
Specifications should therefore be reusable, versioned, machine-checkable, co-owned and independently reviewed by those defining protocol intent and those assessing security. 
The same invariants and reference models should support testing, fuzzing, static
analysis, and proof-assistant verification. 
LLMs may reduce the cost of authoring, translating, and maintaining these artifacts, but cannot establish that they faithfully capture protocol intent.

\point{From AI assistance to checkable assurance}
Our survey separates acceleration from assurance. 
Most respondents report that AI has reduced development or audit effort, but they rate LLMs most highly for bounded tasks such as validating findings, producing proofs of concept, and analyzing conventional application code. 
Trust in LLMs is substantially weaker for circuit- and proof-system-level reasoning.
This suggests that the near-term opportunity is \emph{verified acceleration}: LLMs can propose specifications, invariants, tests, counterexamples, and proof scripts, while independent analyzers, proof kernels, and expert review determine which outputs become assurance evidence.
AI-assisted results should record the provenance and review of generated artifacts, distinguish generated specifications from generated proofs, and identify which outputs were independently checked and continuously re-established in CI.

\point{Threats to validity}
We use a standard methodology~\cite{feldt2010threats} to identify validity threats, which we mitigate where possible.

\textit{Internal.} For \rqref{1}/\rqref{3}, we may miss tools or verification efforts because the ZKP ecosystem evolves quickly. For \rqref{2}, tool outputs may be misclassified; for \rqref{4}/\rqref{5}, survey responses may reflect misunderstandings. We mitigate these threats by searching academic and practitioner sources, preserving raw tool outputs, manually checking ambiguous detections, and piloting the questionnaires.

\textit{Construct.} In \rqref{1}/\rqref{3}, labels such as supported or partial may hide differences in assumptions and trusted components. In \rqref{2}, we measure detection of known vulnerable components, not false positives. For \rqref{4}/\rqref{5}, respondents may interpret tool categories and vulnerability classes differently; we mitigate this with free-text fields and qualitative analysis.

\textit{External.} The \rqref{1}/\rqref{3} results are a snapshot of a fast-moving field. The \rqref{2} results may not generalize beyond Circom and the six tools we evaluated, although Circom is currently the best-supported DSL by automated tools. The \rqref{4}/\rqref{5} survey targets experienced practitioners from high-profile ZKP organizations, improving relevance for high-impact systems but limiting representativeness.

\section{Related Work}
\point{Security Tools and Verification for ZKPs}
Recent studies systematize the security of ZKP systems.
Chaliasos et al.~\cite{chaliasos2024sok} analyze real-world SNARK vulnerabilities across the circuit, frontend, backend, and integration layers, whereas Tang et al.~\cite{tang2024analysis} classify ZKP vulnerabilities. 
Complementary to these works, we study the effectiveness, maturity, and adoption of security tools and verification techniques for ZKPs.

Further, there has been a growing number of security tools for detecting vulnerabilities in ZKP implementations.
These tools range from static analyses~\cite{wen2024practical,kolozyan2025dcm}, SMT-based symbolic techniques~\cite{pailoor2023automated,soureshjani2023halo2,isabel2024civer}, to fuzzing approaches~\cite{chaliasos2025fuzzing,takahashi2025zkfuzz}.
In this work, we conduct a qualitative study of these tools, identify gaps in their coverage, perform the first large-scale evaluation on real-world vulnerabilities, and highlight their limitations relative to the original papers' evaluations.

Due to the significance of ZKPs and the complexity of their implementations, substantial work has focused on formal verification across different parts of the stack.
Existing techniques range from interactive theorem proving~\cite{coglio2023formal,liu2024certifying,chin2021leo}, SMT-based approaches~\cite{ozdemir2025bounded,ozdemir2023satisfiability,hader2024smtlib,isabel2026orchestral}, to automated decision-procedure methods~\cite{jiang2025conscs,yang2026ac4,stephens2025automated}.
We perform the first systematic analysis of all these techniques and highlight major gaps and risks in their current applicability.

In addition to these techniques, there have been efforts in different layers and specialized settings.
These include proof systems~\cite{nguyen2023revisiting,khovratovich2025how},
ZKP compilers~\cite{xiao2025mtzk,hochrainer2025fuzzing}, integrations of ZKPs with smart contracts~\cite{chaliasos2025towards,gomes2025escape,chaliasos2025unaligned}, and fuzzing for specific implementations~\cite{peng2025zkevm,hochrainer2025arguzz}.
In contrast, we focus on generalized security tooling and on the circuit and core implementation layers.
We leave the systematic study and evaluation of more specialized techniques and targets for future work.

%
%
\point{Empirical studies of security tools and surveys}
Empirical studies of security tools and practitioner surveys have been conducted in several domains.
Prior work on smart contracts has surveyed vulnerabilities, verification methods, and analysis tools, and has empirically evaluated security tools against real-world exploits~\cite{chaliasos2024smart,chen2021survey,harz2018towards,kushwaha2022ethereum}.
Other studies have examined developers' expectations of program analysis tools and the usability of security tools~\cite{christakis2016developers}.

\section{Conclusion}
Our study of ZKP security tooling, six automated tools, formal-verification efforts, and 48 practitioners shows that current defenses are useful but incomplete.
Tools find some local underconstraint bugs, but detection drops on whole projects, while formal verification mostly covers isolated constraint-correctness claims rather than end-to-end systems.
Practitioners likewise report human-led workflows and different needs across developers and auditors.
By mapping tool coverage, failures, and practitioner needs, this work gives researchers a concrete agenda for broader, more reliable tooling and gives practitioners a clearer basis for assessing defenses.

\section*{Acknowledgments}

The authors thank Shankara Pailoor for valuable feedback, and the anonymous survey participants for their time and insights. Arman further thanks Carmela Troncoso for her guidance and support throughout his internship at MPI-SP. This work was partially funded by the Ethereum Foundation (EF).

\bibliographystyle{IEEEtran}
\bibliography{references}

\end{document}